\documentclass[%
aps,
prd,
reprint,
amsmath,amssymb,
superscriptaddress,
floatfix,
showpacs
]{revtex4-2}

\usepackage{dcolumn}
\usepackage[mathlines]{lineno}
\usepackage{here}
\usepackage{xcolor}

\usepackage{cprotect}
\usepackage{fancyvrb}

\usepackage{graphicx,hyperref}
\usepackage{orcidlink}

\hypersetup{%
pdfstartview=XYZ,
bookmarksopen,
colorlinks=true,
linkcolor=blue,
citecolor=blue,
urlcolor=blue,
bookmarks=true,
linktocpage=true, 
hyperindex=true
}

\begin{document}

\begin{flushright}
\begin{minipage}{3.8cm}
Belle preprint 2023-01 \\
KEK preprint 2022-48
\end{minipage}
\end{flushright}

\preprint{v.06f}

\title{First measurement of the 
\textbf{\textit{Q}\textsuperscript{2}} 
distribution of \textbf{\textit{X}(3915)} 
single-tag two-photon production}

\noaffiliation
  \author{Y.~Teramoto\,\orcidlink{0000-0002-1738-6697}} 
  \author{S.~Uehara\,\orcidlink{0000-0001-7377-5016}} 
  \author{I.~Adachi\,\orcidlink{0000-0003-2287-0173}} 
  \author{H.~Aihara\,\orcidlink{0000-0002-1907-5964}} 
  \author{S.~Al~Said\,\orcidlink{0000-0002-4895-3869}} 
  \author{D.~M.~Asner\,\orcidlink{0000-0002-1586-5790}} 
  \author{T.~Aushev\,\orcidlink{0000-0002-6347-7055}} 
  \author{R.~Ayad\,\orcidlink{0000-0003-3466-9290}} 
  \author{V.~Babu\,\orcidlink{0000-0003-0419-6912}} 
  \author{Sw.~Banerjee\,\orcidlink{0000-0001-8852-2409}} 
% ...
\collaboration{The Belle Collaboration}

\noaffiliation

\begin{abstract}
We report the first measurement of the $Q^2$ distribution of $X(3915)$
produced by single-tag two-photon interactions. The decay mode used
is $X(3915) \rightarrow J/\psi\omega$. The covered $Q^2$ region is
from 1.5~(GeV/$c$)$^2$ to 10.0~(GeV/$c$)$^2$. 
We observe 
$7.9\pm 3.1({\rm stat.})\pm 1.5({\rm syst.})$ events,
where we expect $4.1\pm 0.7$ events based on the $Q^2=0$ result 
from the no-tag two-photon process, extrapolated to  
higher $Q^2$ region using the $c\bar{c}$ model of Schuler, Berends, 
and van Gulik.
The shape of the distribution is also consistent with this model;
we note that statistical uncertainties are large.
\end{abstract}


\maketitle

\section{\label{introduction}Introduction}

The discovery of $X(3872)$ opened a new era of exotic
hadrons called charmoniumlike states~\cite{X3872-Belle}.
Understanding the nature of this state and of other charmoniumlike 
states, in general, provides an opportunity to study the nonperturbative 
regime of quantum chromodynamics. 
In searching for other charmoniumlike states, 
$X(3915)$ was found by the Belle experiment~\cite{X3915-0,X3915-1} 
and confirmed by the BaBar experiment~\cite{X3915-2,X3915-3},
initially in the study of 
$B^- \rightarrow J/\psi\omega K^-$~\footnote{Charge-conjugate process
will be implicitly included.}
and later in no-tag two-photon interactions.
This state, $X(3915)$, was classified as $\chi_{c0}(3915)$ in the latest 
listing by the Particle Data Group~\cite{PDG-2}, 
but the assignment is not firmly established.
The spin-parity of $X(3915)$ is consistent with $J^P = 0^+$ based on 
the experimental analysis~\cite{X3915-3}; it has a small 
possibility of being $2^+$~\cite{PDG,Zhou}.
If $X(3915)$ is a conventional $c\bar{c}$ state, it should
also decay to $D^{(*)}\bar{D}^{(*)}$ or its charge conjugate. 
In an amplitude analysis of the $B^-\rightarrow K^-D^+D^-$ by the LHCb
experiment, $0^{++}$ and $2^{++}$ states near 3.930~GeV/$c^2$ 
are reported~\cite{LHCb-DD}; they are assigned 
as $\chi_{c0}(3930)$ and $\chi_{c2}(3930)$, respectively.
However, no peaks in $M(D\bar{D}^{(*)})$ have been seen in the studies of 
$B^- \rightarrow D\bar{D}K^-$ and $B^- \rightarrow D\bar{D}^*K^-$
performed by the {\it B}-factories~\footnote{Kinematically 
allowed decay is only $D\bar{D}$ if the mass is 3.920~GeV/$c^2$ 
and $J=0$. Also in $D\bar{D}^*$, no peak is seen.}
\cite{Belle-Ds0D0,DDbar-1,DDbar-3,DDbar-2}. 
Non-$c\bar{c}$ models such as $c\bar{c}s\bar{s}$ models 
or $D_s\bar{D}_s$ molecule models can predict 
such a signature~\cite{Exotics1,Exotics2,Exotics3}.
The $0^{++}$ state(s) reported by LHCb and the {\it B}-factories could 
be different states, namely the 
$\chi_{c0}(2P)$ and non-$c\bar{c}$ state,
respectively.

\begin{figure}[ht]
\centering
  \includegraphics*[width=0.32\textwidth]{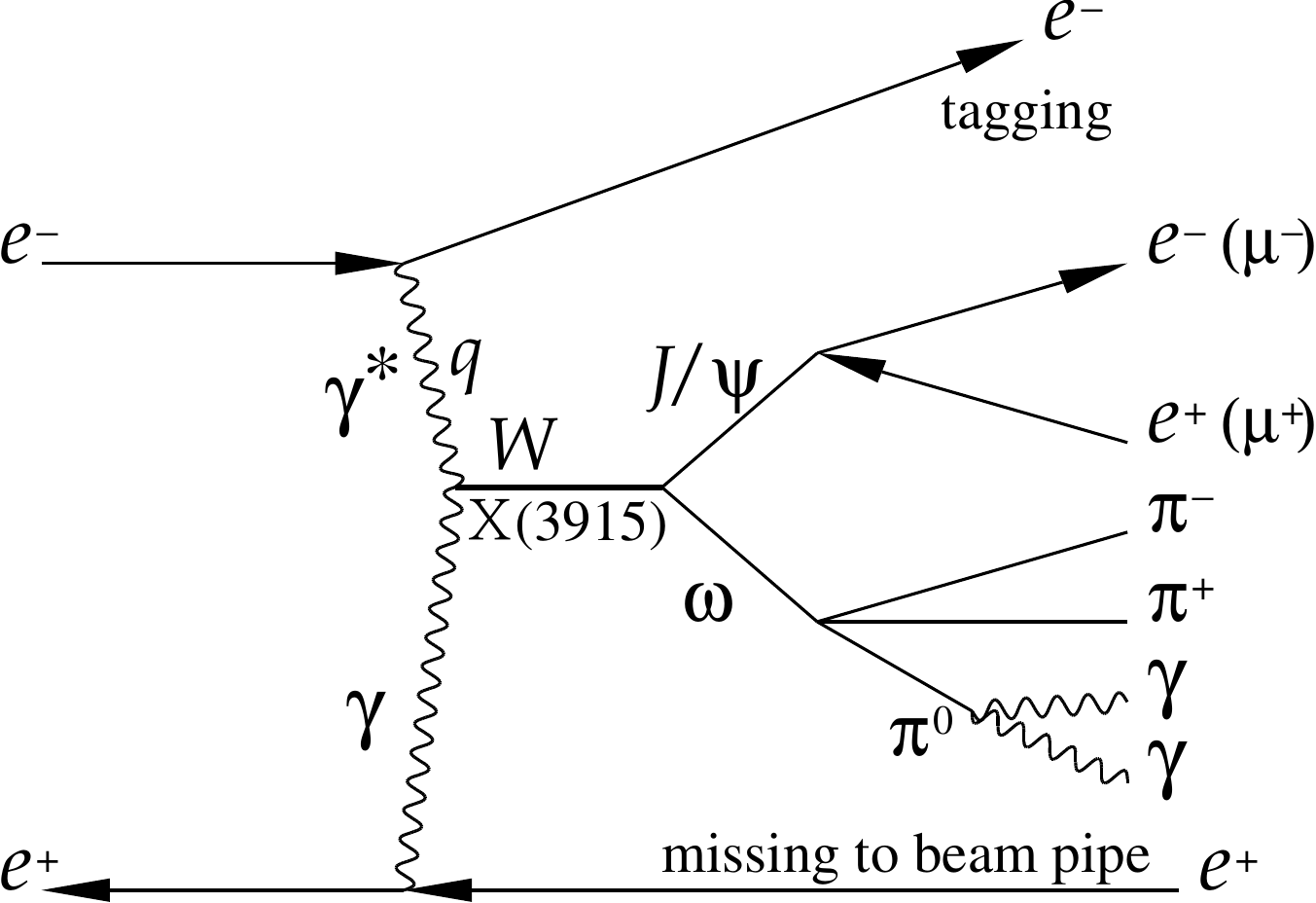}
  \caption{\label{fig:feyn}Single-tag two-photon $X(3915)$ production.
Virtual photon, $\gamma^*$, is produced in the tagging side; 
$q$ is the four-momentum of the $\gamma^*$. 
$W$ is the energy of the two-photon 
system in its rest frame which corresponds to the invariant 
mass of $J/\psi\omega$, $M(J/\psi\omega)$, in this case. 
Tagging is either $e^-$ or $e^+$.
  }
\end{figure}

In this paper, we report on a study of the production of $X(3915)$ 
by highly virtual photons, $\gamma^*$. The reaction used is 
$\gamma^*\gamma\rightarrow X(3915) \rightarrow J/\psi\omega$,
where $\omega$ decays to $\pi^+\pi^-\pi^0$, 
$\pi^0$ decays to two photons and $J/\psi$ decays to
either $e^+e^-$ or $\mu^+\mu^-$, shown in Fig.~\ref{fig:feyn}.
The highly virtual photon is identified by tagging either $e^-$ or $e^+$
in the final state where its partner, $e^+$ or $e^-$, respectively, is missed 
going into the beam pipe. This type of interaction is referred to as a 
``single-tag" two-photon interaction. 
If $X(3915)$ is a non-$c\bar{c}$ state, 
naively it should have a larger
spatial size than $c\bar{c}$. This larger size is predicted 
for charm-molecule models~\cite{Exotics3,Exotics4}. 
In such a case, the production rate should 
decrease steeply at high virtuality. 
To test a deviation from a pure $c\bar{c}$, 
we use a reference $c\bar{c}$ model calculated 
by Schuler, Berends, and van Gulik (SBG)~\cite{SBG}.
In this test, we use the parameter $Q^2$, appearing in its production, 
where $Q^2 (= - q^2)$ is the negative mass-squared 
of the virtual photon; 
$q$ is the four-momentum of the virtual photon.

We will use the term ``electron'' for either the 
electron or the positron.
Quantities calculated in the initial-state $e^+e^-$ 
center-of-mass (c.m.) system are indicated by an asterisk(*).

\section{\label{detector}Detector and Data}

The analysis is based on 825~fb\textsuperscript{-1} of data collected by the
Belle detector operated at the KEKB $e^+e^-$ asymmetric
collider~\cite{KEKB,PTEP-2}. The data were taken at the
$\Upsilon(nS)$ resonances ($n\le 5$) and nearby energies,
9.42~GeV $< \sqrt{s} <$ 11.03~GeV.
The Belle detector was a general-purpose magnetic spectrometer
asymmetrically enclosing the interaction point (IP) 
with almost 4$\pi$ 
solid angle coverage~\cite{Detector,PTEP-1}.
Charged-particle momenta are measured by a silicon vertex detector and a
cylindrical drift chamber (CDC).
Electron and charged-pion identification relies on a combination
of the drift chamber, 
time-of-flight scintillation counters (TOF),
aerogel Cherenkov counters (ACC), 
and electromagnetic calorimeter 
(ECL) made of CsI(Tl) crystals.
Muon identification relies on resistive plate chambers
(RPC) in the iron return yoke. 
Photon detection and energy measurement utilize ECL.

We use Monte Carlo (MC) simulations to set selection
criteria and to derive the reconstruction efficiency.
Signal events, $e^+e^-\rightarrow e^{\pm}(e^{\mp})(\gamma^*\gamma
\rightarrow J/\psi\omega)$, are generated using
TREPSBSS~\cite{TREPSBSS-1,TREPSBSS-2} with a mass 
distribution, centered at $M=3.918$~GeV/$c^2$ and 
width $\Gamma=0.020$~GeV/$c^2$~\cite{PDG}, with constant
transition form factor, $F(Q^2)$=const.
Measured results do not depend on
this setting, as the analysis is performed in bins of $Q^2$.
Decays of the $\omega$ are performed according to the usual
amplitude model~\cprotect\footnote{
\url{https://evtgen.hepforge.org/doc/models.html}, entry EvtOmegaDalitz}. 
Radiative $J/\psi$ decays are
simulated by PHOTOS~\cite{PHOTOS-1,PHOTOS-2}.
Detector response is simulated employing 
GEANT3~\cite{Geant}.

\section{\label{PID}Particle Identification}

Final-state particles in this reaction are 
$\ell^+\ell^-\pi^+\pi^-\gamma\gamma$ where $\ell^+\ell^-$ is either 
an electron pair or a muon pair.

Electrons are identified using a combination of five 
discriminants:
$E/p$, where $E$ is the energy measured by ECL
and $p$ is the momentum of the particle, then, transverse shower shape 
in ECL, position matchings between the energy cluster and the extrapolated 
track at ECL, ionization loss in CDC, and light yield in ACC. 
For these, probability density functions are derived 
and likelihoods, $L_i$'s, are calculated, where $i$'s stand 
for the discriminants. Electron likelihood ratio, ${\cal L}_e$, is 
obtained by $\Pi_i L_i^{\rm electron}/(\Pi_i L_i^{\rm electron} + 
\Pi_i L_i^{\rm nonelectrons})$~\cite{eID}.

Muons are identified using a combination of two measurements:
penetration depth in RPC, and deviations of hit-positions in RPC 
from the extrapolated track. 
From these, the muon likelihood ratio, ${\cal L}_{\mu}$, is 
obtained by $P_{\mu}/(P_{\mu}+P_{\pi}+P_K)$, where $P_{\mu}$, 
$P_{\pi}$, and $P_K$ are probabilities for muon, pion, and kaon, 
respectively~\cite{muonID}.

Charged pions and kaons are identified using the combination of
three measurements: ionization loss in CDC, time-of-flight by TOF, 
light-yield in ACC. From these, the pion likelihood ratio, 
${\cal L}_{\pi}$, is calculated by $P_{\pi}/(P_K + P_{\pi})$ where $P_{\pi}$ 
and $P_K$ are pion and kaon probabilities, respectively~\cite{PID}.

Photons are identified by position isolations
between the energy cluster and the extrapolated track at ECL.

\section{\label{selection}Event Selection}

Event-selection criteria share the ones in our previous 
publication~\cite{X3872-our}.
We select events with five charged tracks coming from the IP
since one final-state electron goes into the beam pipe
and stays undetected.
Each track has to have $p_T > 0.1$~GeV/$c$, with two or more
having $p_T > 0.4$~GeV/$c$, where $p_T$ is the transverse
momentum with respect to the $e^+$ beam direction.
Total charge has to be $\pm 1$.

$J/\psi$ candidates are reconstructed by their decays to lepton pairs:
$e^+e^-$ or $\mu^+\mu^-$. Electrons are identified by requiring 
${\cal L}_e$ to be greater than 0.66 having 90\% efficiency.
Similarly, muons are identified by requiring ${\cal L}_{\mu}$
to be greater than 0.66 having 80\% efficiency.
We require the invariant mass of the lepton
pair to be in the range [3.047 GeV/$c^2$; 3.147 GeV/$c^2$].
In the calculation of the invariant mass of an $e^+e^-$ pair, we
include the four-momenta of radiated photons if the photons have
energies less than 0.2~GeV and polar angles, relative to 
the electron direction at the IP, less than 0.04~rad.

For the tagging electron, a charged track has to satisfy 
${\cal L}_e$ greater than 0.95 or $E/p$ greater 
than 0.87.
In addition, we require $p > 1.0$~GeV/$c$ and $p_T > 0.4$~GeV/$c$.
In the calculation of $p$, four-momenta of radiated photons are
included using the same requirements as for the electrons 
from $J/\psi$ decays.

Charged pions are identified by satisfying its
${\cal L}_{\pi}$ be greater than 0.2, 
${\cal L}_{\mu}$ less than 0.9, 
${\cal L}_e$ less than 0.6
and $E/p$ less than 0.8, having 90\% efficiency.

Neutral pions are reconstructed from their decay photons,
where the photons are identified as energy clusters 
in the electromagnetic calorimeter and isolated from charged tracks.
These photons have to fulfill the requirements 
$E_{\gamma{\rm H}}<-7 E_{\gamma{\rm L}}
+ 0.54$~GeV and $E_{\gamma{\rm H}}>0.12$~GeV, where $E_{\gamma{\rm H}}$ is 
the energy of the higher-energy photon, and 
$E_{\gamma{\rm L}}$ is the energy of the lower-energy photon, both in GeV.
Neutral-pion candidates have to satisfy $\chi^2 < 9$ 
for their mass-constraint fit. 
If there is only one $\pi^0$ candidate with $p_T > 0.1$~GeV/$c$,
we accept the one as $\pi^0$. If there is no such $\pi^0$, 
but there are one or more
$\pi^0$ candidates with $p_T < 0.1$~GeV/$c$, we calculate the invariant mass,
$M(\pi^+\pi^-\pi^0)$, for each $\pi^0$ candidate. If there is only one
candidate having its $M(\pi^+\pi^-\pi^0)$ in the $\omega$-mass region 
[0.7326~GeV/$c^2$; 0.8226~GeV/$c^2$], we accept the one as $\pi^0$.
If more than one candidate satisfy the $\omega$-mass condition,
we accept the one with the smallest mass-constraint fit $\chi^2$ as $\pi^0$.
If there are more than one $\pi^0$ candidate with $p_T > 0.1$~GeV/$c$,
we test the $\omega$-mass condition for each $\pi^0$ candidate. 
If there is only one candidate that satisfies the $\omega$-mass condition,
we accept it as $\pi^0$. If more than one such candidate exist,
we accept the one with the smallest mass-constraint fit $\chi^2$ as $\pi^0$.

Events should not have $e^+e^-$ pairs from $\gamma\rightarrow e^+e^-$.
Therefore, we discard the event if the invariant mass of the pair of 
any oppositely charged tracks is less than 0.18~GeV/$c^2$,
calculated assuming them as electrons. 
We require that the event has no photon with energy above 0.4~GeV.
Events must have one $\omega$ identified by the $\omega$-mass condition.

The tagging electron and the rest of the particles should 
be back-to-back, projected in the plane perpendicular to the 
$e^+$ beam axis. For this, we require $||\phi({\rm tag}) 
- \phi({\rm rest \ combined})| - \pi | < 0.15$~rad, where $\phi$
is the azimuthal angle about the $e^+$ beam axis.

A missing momentum arises from the momentum of the final-state
electron that goes undetected into the beam pipe. We require the 
missing-momentum projection in the $e^-$ beam direction in the c.m. system 
to be less than $-0.2$~GeV/$c$ for $e^-$-tagging events and greater than
0.2~GeV/$c$ for $e^+$-tagging events.
The upper limit on the $Q^2$ of untagged photons 
is estimated to be 0.1~(GeV/$c$)\textsuperscript{2}. 

\begin{figure}[ht]
\centering
  \includegraphics*[width=0.48\textwidth]{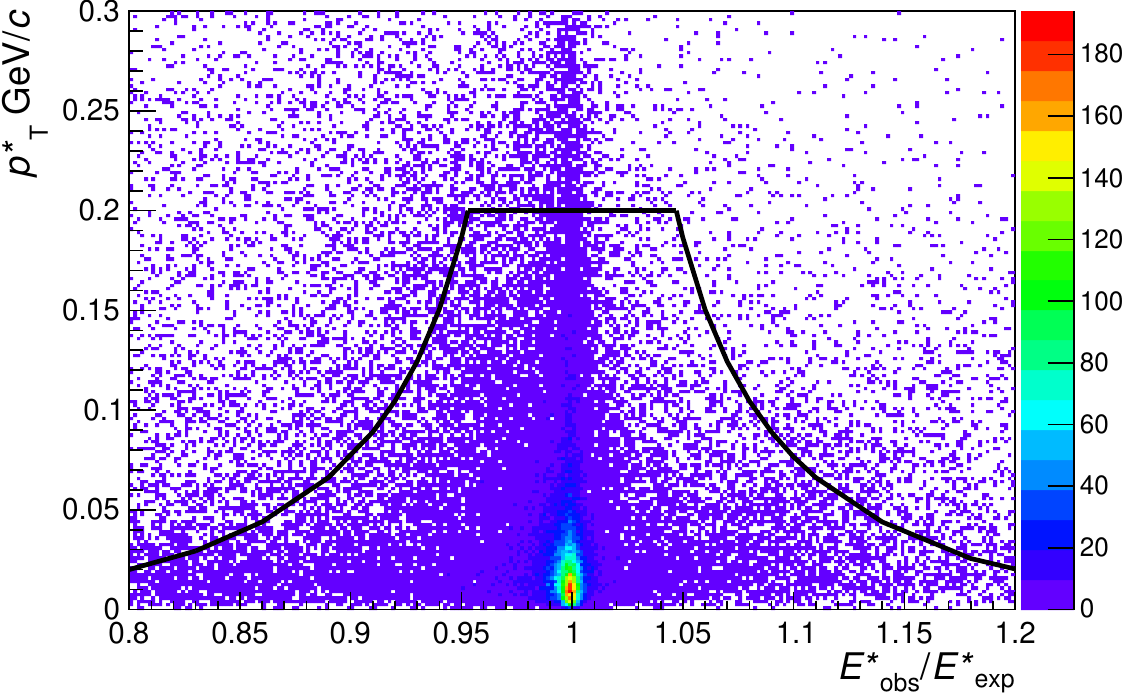}
  \caption{\label{fig:hyper}$p^*_T$ vs.~$E^*_{\rm obs}/E^*_{\rm exp}$
distribution (MC events). The (black) line shows the selection criteria
applied to $p^*_T$ and $E^*_{\rm obs}/E^*_{\rm exp}$; events below
the line are accepted.
  }
\end{figure}

The total visible transverse momentum of the event, $p^*_T$, should
be less than 0.2~GeV/$c$. Measured energy of 
the $J/\psi\pi^+\pi^-\pi^0$ system, $E^*_{\rm obs}$,
should be equal to the expected energy, $E^*_{\rm exp}$, 
calculated from the momentum of the tagging electron and the direction
and invariant mass of the $J/\psi\pi^+\pi^-\pi^0$ system.
Since energy and $p^*_T$ are correlated, we impose a
two-dimensional criterion
\begin{equation}\label{eq:hyper}
\begin{array}{l}
{\displaystyle
  (p^*_T + 0.04~{\rm GeV}/c)\left(\frac{|E^*_{\rm obs}-E^*_{\rm exp}|}
{E^*_{\rm exp}}+0.003\right)
} \\
\multicolumn{1}{r}{<0.012~{\rm GeV}/c.}
\end{array}
\end{equation}
Figure~\ref{fig:hyper} shows the $p^*_T$ vs.~$E^*_{\rm obs}/E^*_{\rm exp}$
distribution from MC events with the selection criteria. 

A non-signal event imitates $X(3915)$ if a $\psi(2S)$ is produced by 
a virtual photon from internal bremsstrahlung 
and if it accompanies either a $\pi^0$ or a fake $\pi^0$ 
and also the $\pi^+\pi^-\pi^0$ combination satisfies the 
$\omega$-mass condition. 
To suppress this background, we reject the event having the invariant 
mass of $J/\psi\pi^+\pi^-$ in the $\psi(2S)$ 
window [3.6806~GeV/$c^2$; 3.6914~GeV/$c^2$]. This window is defined as 
$\pm 2\sigma$ of the $\psi(2S)$ mass resolution.
The mass resolutions of $\psi(2S)$(= $J/\psi\pi^+\pi^-$) 
and $X(3915)$(= $J/\psi\pi^+\pi^-\pi^0$) 
are approximately 2.7~MeV/$c^2$.

\section{\label{result}Results}
\subsection{\label{signal}Signals and backgrounds}

\begin{figure}[ht]
\centering
  \includegraphics*[width=0.47\textwidth]{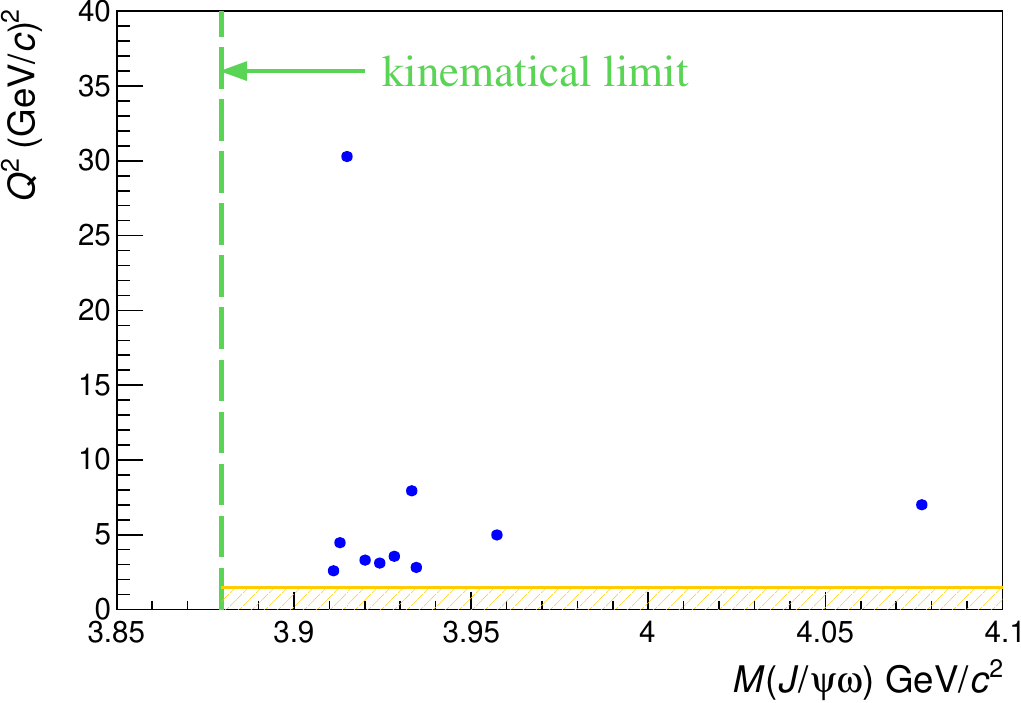}
  \caption{\label{fig:open-2D}$Q^2$ vs.~$M(J/\psi\omega)$
distribution from data. The dashed (green) line 
indicates the kinematical limit: 3.8795~GeV/$c^2$. 
The hatched (orange) region 
has detection efficiency below 15\% of its 
plateau value as explained in the text.
  }
\end{figure}

Figure~\ref{fig:open-2D} shows the $Q^2$ vs.~$M(J/\psi\omega)$
distribution from the selected data. Here, $Q^2$ is calculated by 
$Q^2 = 2(p_{\rm beam}p_{\rm tag} - m_e^2)$, where $p_{\rm beam}$ and 
$p_{\rm tag}$ are the four-momenta of the beam $e^{\pm}$ and tagging $e^{\pm}$,
respectively, and $m_e$ is the electron mass. 
The events fall into three classes: a cluster in the
$X(3915)$ mass region with $Q^2$ less than 10~(GeV/$c$)\textsuperscript{2},
a high $Q^2$ event at $Q^2\approx 30$~(GeV/$c$)\textsuperscript{2}, 
and a high $M$ event at $M\approx 4.08$~GeV/$c^2$.
In the small $Q^2$ region, the detection efficiency 
diminishes due to the electron tagging condition [see Appendix, 
Fig.~\ref{fig:eff}]. This region,
$Q^2 < 1.5$~(GeV/$c$)\textsuperscript{2}, is hatched in 
Fig.~\ref{fig:open-2D}, where the detection efficiency falls 
below 15\% of its plateau value.

\begin{figure}[ht]
  \centering
  \includegraphics*[width=0.47\textwidth]{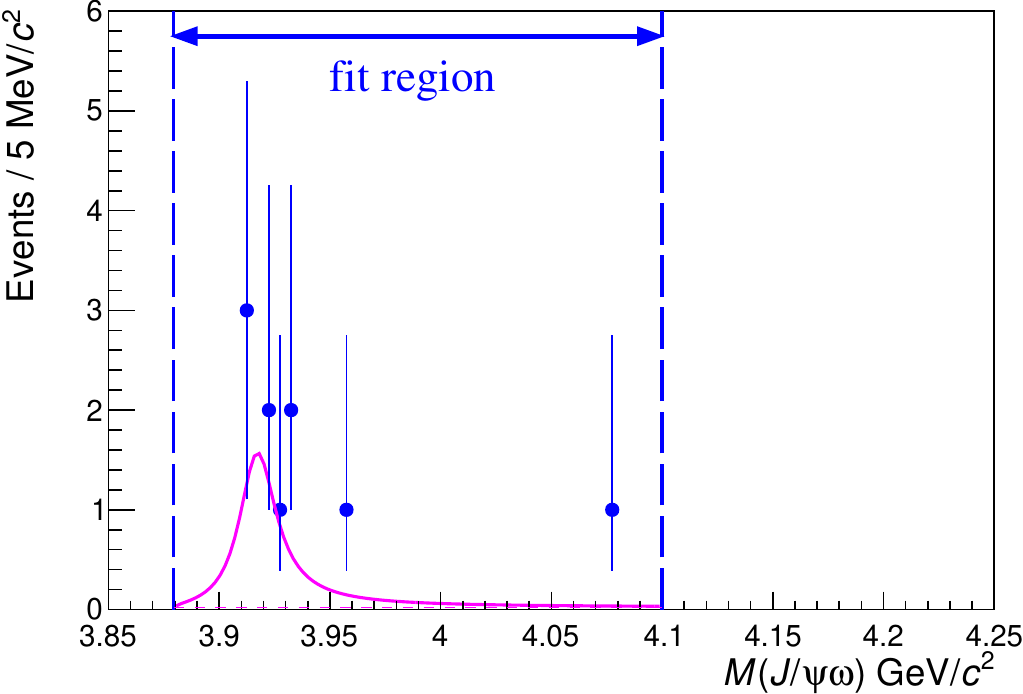}
  \caption{\label{fig:BW_lin}$M(J/\psi\omega)$ distribution 
with the Breit--Wigner + flat function fit.
Abscissa is $M(J/\psi\omega)$ in GeV/$c^2$. Ordinate is the
number of events per 5~MeV/$c^2$. 
Solid (magenta) curve shows the result of the fit. 
Horizontal dashed (magenta) line shows the flat component.
Vertical dashed (blue) lines indicate the fit 
region.}
\end{figure}

To derive the numbers of signal and background events, 
we fit a combination of the threshold-corrected 
relativistic Breit--Wigner (BW) function and
a constant to the  $M(J/\psi\omega)$ distribution. 
The threshold-corrected BW function, $f_{\rm BW}(W)$, is

\begin{equation}\label{eq:std-BW}
{\displaystyle
f_{\rm BW}(W) = 
\frac{\alpha M^2\Gamma'}{(W^2-M^2)^2+M^2\Gamma'^2}
}
\end{equation}
where $M$ is the resonance mass, 
$\alpha$ is a dimensionless normalization factor,
and $\Gamma'$ is the threshold-corrected resonance width
defined by
\begin{equation}
{\displaystyle
\Gamma' = \Gamma \cdot \frac{\rho(W)}{\rho(M)}
}
\end{equation}
where $\Gamma$ is the resonance width, 
$\rho(W)$ is the phase space factor for $W$, which is
\begin{equation}\label{eq:phase-space}
{\displaystyle
\rho(W) = \frac{1}{16\pi}\frac{\lambda^{1/2}(W^2, m_J^2, m_{\omega}^2)}{W^2}
}
\end{equation}
and $\lambda$ is the K{\"a}ll{\'e}n function~\cite{PDG-2,Kallen}. 
It is defined as
\begin{equation}
\begin{array}{l}
{\displaystyle
\lambda^{1/2}(W^2, m_{J/\psi}^2, m_{\omega}^2) 
} \\
{\displaystyle
  = \sqrt{(m_{J/\psi}^2 + m_{\omega}^2 - W^2)^2
    - 4m_{J/\psi}^2m_{\omega}^2}
}
\end{array}
\end{equation}
where $m_{J/\psi}$ is the mass of $J/\psi$(= 3.0969~GeV/$c^2$) and
$m_{\omega}$ that of $\omega$(= 0.78265~GeV/$c^2$)~\cite{PDG}.
In the fit, we set $M=3.918$~GeV/$c^2$, 
$\Gamma = 0.020$~GeV/$c^2$~\cite{PDG}, and $\alpha = 2/\pi$, 
with the fit function (modified BW combined with a flat component)
\begin{equation}\label{eq:BW-fit}
{\displaystyle
f_{\rm BW + flat} = a_{\rm BW}\cdot f_{\rm BW} + a_{\rm flat},
}
\end{equation}
where the fit parameters $a_{\rm BW}$ and $a_{\rm flat}$ are 
the magnitudes of the BW and the flat component, respectively. 
We ignore a possible distortion of the fit distribution 
due to the energy dependence of the detection sensitivity,
because the effect is small. 
Energy dependence of the detection sensitivity for 
$J/\psi\omega$, which is defined by the production of 
detection efficiency times luminosity function, is estimated 
as $0.1\Delta W$~\%, where $\Delta W$ is in the MeV unit.
We use the ROOT/MINUIT implementation of the binned maximum-likelihood 
method with a 5~MeV/$c^2$ bin width and perform the fit in the
$M(J/\psi\omega)$ range of [3.880 GeV/$c^2$; 4.100 GeV/$c^2$].
The units of $f_{\rm BW + flat}$ and $f_{\rm BW}$ are events/(5~MeV/$c^2$) 
and (GeV/$c^2$)$^{-1}$, respectively.
The result of the fit is shown in Fig.~\ref{fig:BW_lin}.

The obtained parameters are 
$a_{\rm BW} = 0.049\pm 0.018$~GeV/$c^2$/(5~MeV/$c^2$)
and $a_{\rm flat} = 0.022\pm 0.035$~/(5~MeV/$c^2$).
The number of signal events is $n_{\rm sig} = 9.0\pm 3.2$,
obtained by integrating $f_{\rm BW}$ with $a_{\rm BW}$ over the fit 
region [3.8795~GeV/$c^2$; 4.1000~GeV/$c^2$].
The number of background events is $n_{\rm bg}^{\rm fit} = 0.3 \pm 0.4$,
calculated for the $X(3915)$ band, which we define 60~MeV/$c^2$. 
It is obtained by multiplying $a_{\rm flat}$ by 
the ratio of the $X(3915)$ band width, 60~MeV/$c^2$, 
to the bin width, 5~MeV/$c^2$. 

\begin{figure}[ht]
  \centering
  \includegraphics*[width=0.47\textwidth]{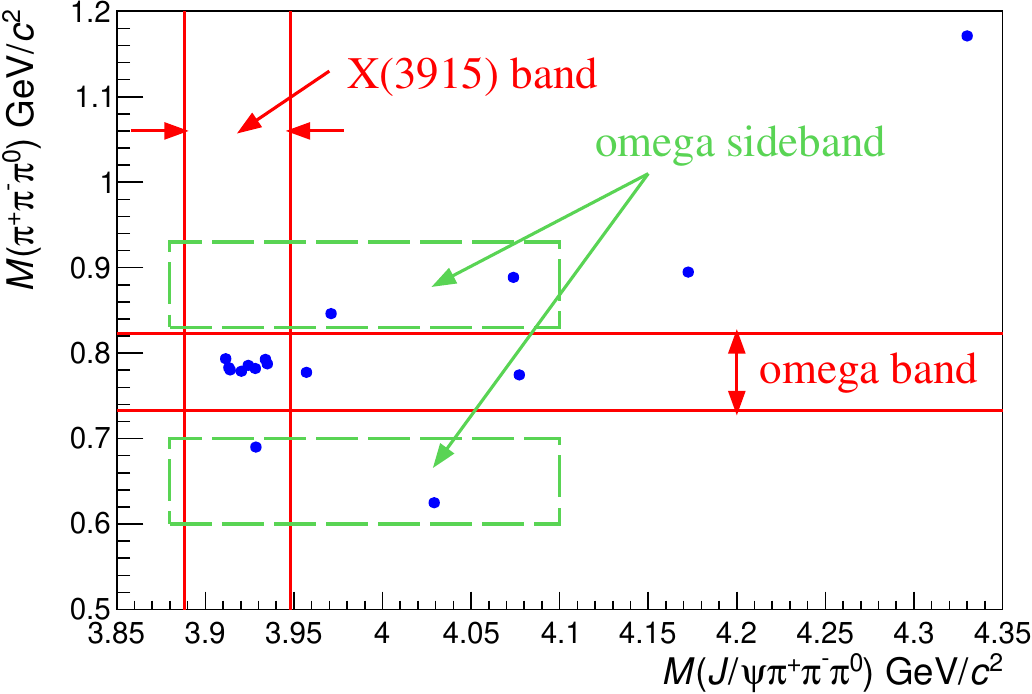}
    \caption{\label{fig:Mx-Mhad}$M(\pi^+\pi^-\pi^0)$ 
vs.~$M(J/\psi\pi^+\pi^-\pi^0)$ distribution to estimate the background rate
using $\omega$ sideband events. 
Dashed (green) rectangles show the $\omega$ sidebands. Horizontal
(red) line pair shows the $\omega$ signal band. Vertical (red) line-pair 
shows the $X(3915)$ signal band.
    }
\end{figure}

To confirm the number of background events, it is 
also derived using the number of events in the $\omega$ sidebands. 
Figure~\ref{fig:Mx-Mhad} shows the $M(\pi^+\pi^-\pi^0)$ 
vs.~$M(J/\psi\pi^+\pi^-\pi^0)$ distribution.
The sideband regions are set as two rectangles with heights
$M(\pi^+\pi^-\pi^0)$ in [0.60,0.70]~GeV/$c^2$ and 
[0.83,0.93]~GeV/$c^2$ and the width $M(J/\psi\pi^+\pi^-\pi^0)$ 
in [3.88,4.10]~GeV/$c^2$. There are in total four events in the 
$\omega$ sideband rectangles.
For the signal region, a rectangle of 0.080~GeV/$c^2$ high in 
$M(\pi^+\pi^-\pi^0)$ and 0.060~GeV/$c^2$ wide in 
$M(J/\psi\pi^+\pi^-\pi^0)$ is used.
From this, the obtained number of background events is 
$n_{\rm bg}^{\omega}= 0.4 \pm 0.3$.
As $n_{\rm bg}^{\omega}$ is calculated using non-$\omega$
events while $n_{\rm bg}^{\rm fit}$ is obtained from identified $\omega$
events, the contents in the samples are different. Nevertheless,
the results from the two methods are approximately the same. 
We use a conservative number: $n_{\rm bg}=0.4 \pm 0.4$.
The resulting signal significance for nine observed events is then 5.6$\sigma$.

The measured number of signals is compared to 
the expectation, $n_{\rm sig}^{\rm exp}$,
derived from the existing no-tag two-photon measurement~\cite{PDG,PDG-2}.
For this, we use the spin-parity of 
$X(3915)$ as $J^P = 0^+$ and use
Eqs.~(\ref{eq:SBG-2}) and (\ref{eq:SBG-0+}) from the SBG model
to extrapolate the $Q^2=0$ value, 
$\Gamma_{\gamma\gamma}(0){\cal B}(X\rightarrow J/\psi\omega)=(54\pm 9)$~eV/$c^2$~\cite{PDG,PDG-2},
to higher $Q^2$, where $\Gamma_{\gamma\gamma}(0)$ 
is the $\gamma\gamma$ decay width of $X(3915)$ at 
$Q^2=0$ and ${\cal B}(X\rightarrow J/\psi\omega)$ is the 
branching fraction of $X(3915)$ decaying to $J/\psi\omega$. 
The result is $n_{\rm sig}^{\rm exp} = 4.1\pm 0.7$.
For a different prediction, if we assume the spin-parity 
as $J^P=2^+$, the expectation is $7.5\pm 1.3$ events using 
$\Gamma_{\gamma\gamma}(0){\cal B}(X\rightarrow J/\psi\omega)=16$~eV 
in Ref.~\cite{X3915-1} with the $J=2$ SBG model, Eq.~(\ref{eq:SBG-2+}),
assuming $\epsilon=1.0$. 

\subsection{\textbf{\textit{Q}\textsuperscript{2}} distribution}

To determine the $Q^2$ distribution, we 
must first determine the treatment of the two outlier events
in Fig.~\ref{fig:open-2D}. The event at $M\approx 4.08$~GeV/$c^2$
is excluded because it is far outside the $X(3915)$ region.
The event at $Q^2 \approx 30$~(GeV/$c$)\textsuperscript{2}
is discussed in the following.

\begin{figure}[ht]
  \centering
  \includegraphics*[width=0.47\textwidth]{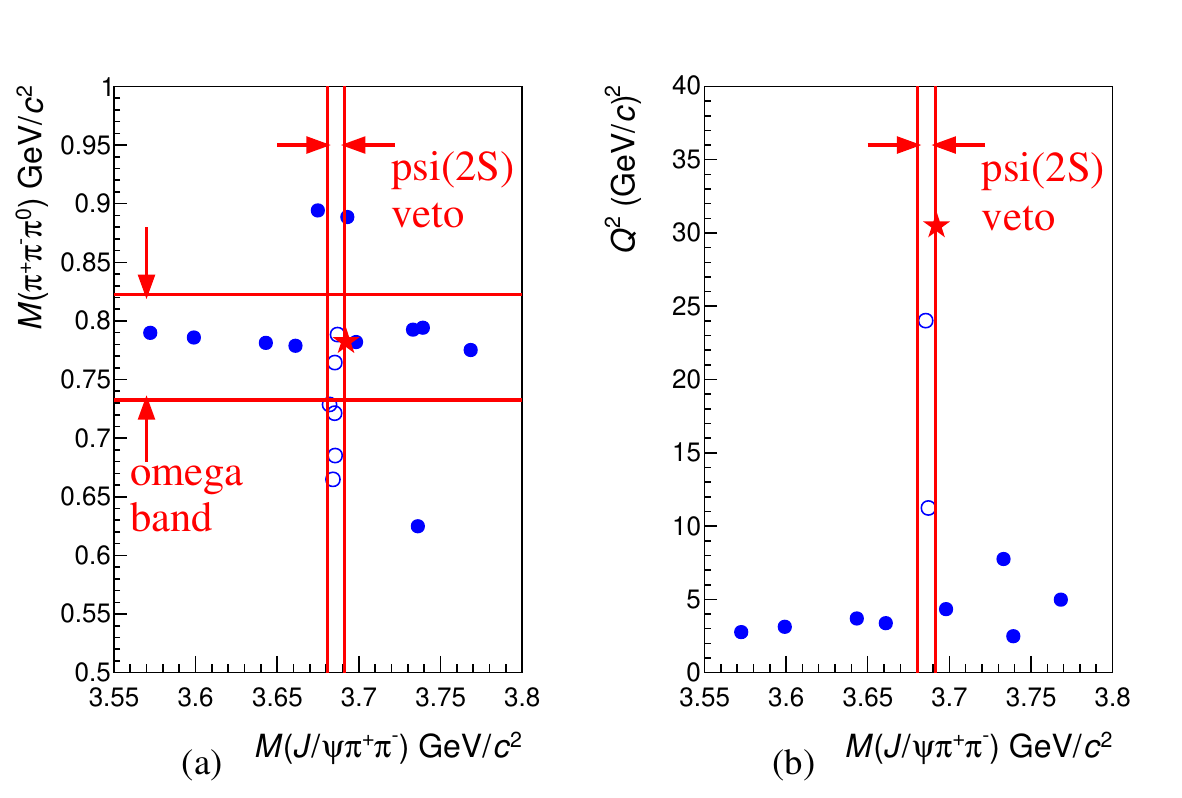}
    \caption{\label{fig:psi2S-pT-omega}(a) $M(\pi^+\pi^-\pi^0)$ 
vs.~$M(J/\psi\pi^+\pi^-)$ and (b) $Q^2$ 
vs.~$M(J/\psi\pi^+\pi^-)$ distributions.
Star (red): high-$Q^2$ event. Open circles (blue): $\psi(2S)$-vetoed
events. Closed circles (blue): $X(3915)$ candidates.
(a) neither  $\omega$ selection nor $\psi(2S)$ veto are applied.
(b) events pass the $\omega$ selection but no $\psi(2S)$ veto is applied.
Vertical red lines indicate the $\psi(2S)$-veto window; 
the horizontal red lines in panel (a) are the $\omega$ signal band.
}
\end{figure}

Figure~\ref{fig:psi2S-pT-omega}(a) shows the 
$M(\pi^+\pi^-\pi^0)$ vs.~$M(J/\psi\pi^+\pi^-)$ distribution,
applying neither the $\omega$ selection
nor the $\psi(2S)$ veto. 
The high-$Q^2$ event is located at 0.9~MeV/$c^2$
above the upper boundary of the $\psi(2S)$ veto. There are six 
events in the $\psi(2S)$ veto. Of the six events, two pass the 
$\omega$ selection.
Figure~\ref{fig:psi2S-pT-omega}(b) shows the $Q^2$ 
vs.~$M(J/\psi\pi^+\pi^-)$ distribution. 
The two $\psi(2S)$-vetoed events, in addition to the high-$Q^2$ event, 
have high $Q^2$s: $Q^2 > 10$~(GeV/$c$)$^2$. 
All the other events that pass the $\psi(2S)$ veto have a lower $Q^2$, i.e.,
$Q^2 < 10$~(GeV/$c$)$^2$.
From this we conclude that $\psi(2S)$-vetoed events have
significantly higher $Q^2$ than the $X(3915)$ events.

As for the possibility of the high-$Q^2$ event being 
an $X(3915)$ signal, the Belle experiment had little sensitivity 
to measure single-tag two-photon events with $Q^2$ around 30~(GeV/$c$)$^2$
as detailed in the Appendix (see, e.g., Fig.~\ref{fig:eff-lum}). 
Hence, it is improbable for the high-$Q^2$ event to be a 
single-tag two-photon event.

To estimate the probability of having one $\psi(2S)\pi^0$ event in the
region adjacent to the $\psi(2S)$ veto window, where the high-$Q^2$ event
is located, we estimate the probability of $\psi(2S)$ events escaping the
veto and having a $\pi^0$.
For this, we employ the data sample used in the 
$X(3872)$ search and examine the $M(J/\psi\pi^+\pi^-)$ 
distribution~\cite{X3872-our}. There are 231 events
in the $\psi(2S)$-veto window of $\pm 5.4$~MeV/$c^2$ used
in the current study. There are 12 events in the 2.7~MeV bin, 
adjacent to the upper boundary of the veto, where the high-$Q^2$
event is located. If we normalize the number of 
events in the veto window
to six that we observe as $\psi(2S)\pi^0$s in this study, 
those 12 events correspond to 0.31 events/bin, or 0.11 events/MeV. 
As seen in Fig.~\ref{fig:psi2S-pT-omega}(a), two out of six 
events are inside the $\omega$ region. Hence, the expected number 
of veto leaks is 0.04~events/MeV. Then, by assuming the width 
of the leak region as 2~MeV and the uncertainty in the number of events 
as 0.1~events, the number of expected events is estimated 
to be $0.1\pm 0.1$~events. 
Significance of that number exceeding one event is 1.5~$\sigma$,
or 7\%. 

A possible way of producing $\psi(2S)\pi^0$ is by a virtual photon, 
radiated by internal bremsstrahlung from $e^-$ or $e^+$, 
similar to the case of $\psi(2S)$ production. 
However, there are suppressions to the $\psi(2S)\pi^0$ production compared 
to $\psi(2S)$. The $\psi(2S)$s are produced as resonances, 
but the $\psi(2S)\pi^0$s are not. 
In order to be $J^P = 1^-$, the $\psi(2S)\pi^0$ has to be in a $P$-wave.
In addition, $\psi(2S)\pi^0$ is an isospin one state. Thus, further 
suppressions are expected.

In the arguments up to this point, we assume the $\pi^0$s as real. 
However, the reconstructed $\pi^0$s can be fake. 
Using MC events, we observe that 13\% of $\pi^0$s, found in the 
$X(3915)$ candidates, are fake.
This number is considered a lower limit as we found that the abundance of
low-$p_T$ $\pi^0$s is higher in real data than in MC.
Thus, the fraction of fake $\pi^0$s is higher at low $p_T$ than at
high $p_T$.
The observed $\psi(2S)\pi^0/\psi(2S)$ is 6/231, where the $\pi^0$s
are either real or fake.
In summary, it is plausible that the high-$Q^2$ event is
a $\psi(2S)\pi^0$ background.

If we remove the high-$Q^2$ event from the $f_{\rm BW + flat} $ fit,
the result is $a_{\rm BW} = 0.043\pm 0.017$~GeV/$c^2$/(5~MeV/$c^2$) and 
$a_{\rm flat} = 0.025\pm 0.036$~/(5~MeV/$c^2$).
From that, we obtain $n_{\rm sig} = 7.9^{+3.1}_{-3.0}$. 
As a note, the significance for eight events is 5.2$\sigma$.

\begin{figure}[t]
  \centering
  \includegraphics*[width=0.47\textwidth]{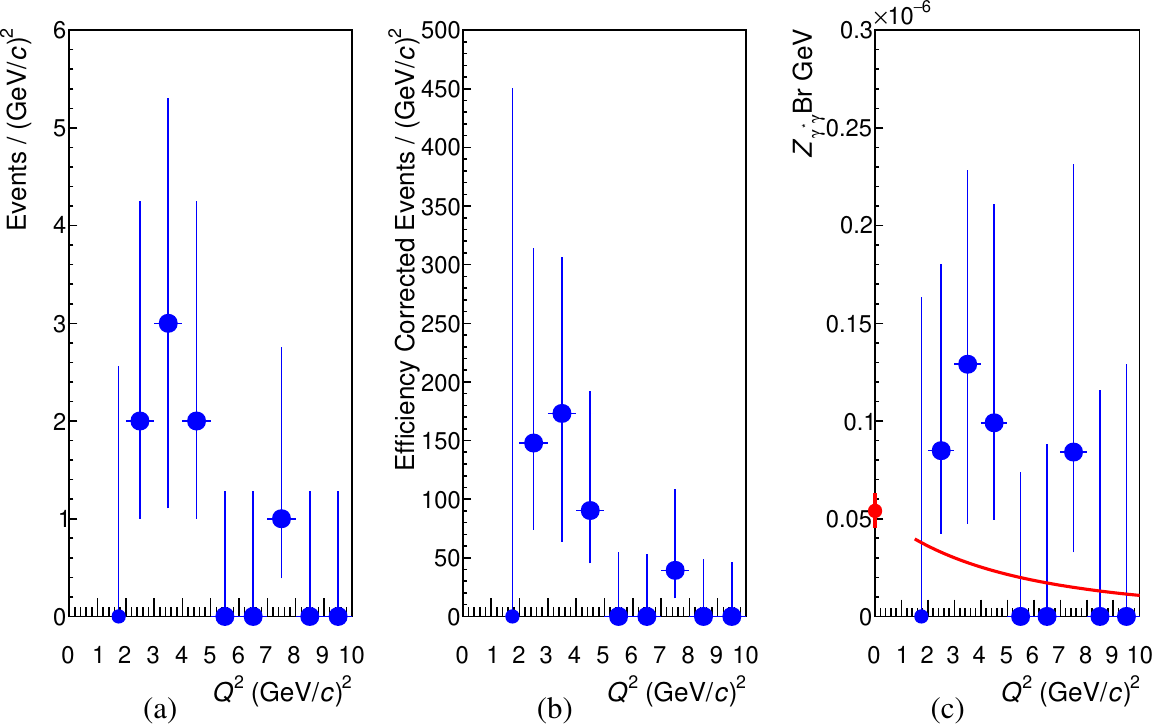}
  \caption{\label{fig:Q2-dist}Measured $Q^2$ distributions:
(a) the number of events per (GeV/$c$)$^2$, 
(b) efficiency corrected number of events per (GeV/$c$)$^2$, and 
(c) $Z_{\gamma^*\gamma}{\cal B}(X\rightarrow J/\psi\omega)$. 
Bin widths of all data are 1~(GeV/$c$)$^2$ except
the smallest $Q^2$ bins whose bin width is 0.5~(GeV/$c$)$^2$.
The solid (red) curve shows the SBG prediction 
based on the data of the no-tag two-photon measurement,
$\Gamma_{\gamma\gamma}(0){\cal B}(X\rightarrow J/\psi\omega)=54$~eV/$c^2$, 
shown as a small (red) circle. 
  }
\end{figure}

In the low-$Q^2$ region, there are eight events in the
$M(J/\psi\pi^+\pi^-\pi^0)$ range [3.911~GeV/$c^2$; 3.958~GeV/$c^2$]. 
In the following, we will study the $Q^2$ structure 
of $X(3915)$ using these eight events, excluding the high-$Q^2$ event and 
the high-$M$ event, which are considered as backgrounds.
Figure~\ref{fig:Q2-dist} shows the $Q^2$ distributions for 
three quantities: 
the number of events, efficiency corrected number of events, and 
$Z_{\gamma^*\gamma}{\cal B}(X\rightarrow J/\psi\omega)$,
where $Z_{\gamma^*\gamma}$ is a $Q^2$-dependent decay function 
defined by Eq.~(\ref{eq:Zgamma}). 
The $Z_{\gamma^*\gamma}{\cal B}(X\rightarrow J/\psi\omega)$ distribution
is obtained by multiplying the event distribution by a correction 
function further detailed in the Appendix [see Eq.~(\ref{eq:C})].
The integrated yield of the 
$Z_{\gamma^*\gamma}{\cal B}(X\rightarrow J/\psi\omega)$ 
distribution in the $Q^2$ range of 1.5~(GeV/$c$)$^2$ to 10.0~(GeV/$c$)$^2$ 
is $1.9 \pm 0.9$ times the expectation from the no-tag measurement, 
$\Gamma_{\gamma\gamma}(0){\cal B}(X(3915)\rightarrow J/\psi\omega)=54\pm 9$~eV~\cite{PDG,PDG-2}, 
combined with its extrapolation to the higher-$Q^2$ region 
using Eq.~(\ref{eq:SBG-0+}).
The averages, $\langle Q^2 \rangle$, 
and the root-mean-squared (rms) values of $Q^2$,
$\sqrt{\langle(Q^2 - \langle Q^2\rangle)^2 \rangle}$, for the 
$Z_{\gamma^*\gamma}{\cal B}(X\rightarrow J/\psi\omega)$ distribution 
are listed in Table~\ref{tab:Q2-dist}, both for the measurement and for 
the SBG model [see Eq.~(\ref{eq:SBG-0+}) of the Appendix].
They are obtained from the same range in $Q^2$ as above, i.e., 
$1.5~(\text{GeV}/c)^{2} \le Q^2 \le 10.0~(\text{GeV}/c)^{2}$.
The measured average $Q^2$, $4.5\pm 0.7$~(GeV/$c$)$^2$, agrees with 
the theoretical prediction, 4.8~(GeV/$c$)$^2$. Their difference 
is approximately 10\% of the rms widths of their distributions, 
which are $1.9\pm 0.8$~(GeV/$c$)$^2$ vs. 2.4~(GeV/$c$)$^2$.
The resolution in $Q^2$  is about 0.03~(GeV/$c$)$^2$ depending
on the tag-electron's scattering angle. 
Hence, the measurement is consistent with the prediction both in 
the averages and the rms values of $Q^2$.
In summary, the measured $Q^2$ distribution does not show a 
significant shift to lower $Q^2$; it agrees with 
the SBG $c\bar{c}$ model. 

\begin{table}[t]
  \centering
  \caption{\label{tab:Q2-dist}Comparison of the measurement and the SBG model
prediction~\cite{SBG} for the $Q^2$ distribution of
$Z_{\gamma^*\gamma}{\cal B}(X\rightarrow J/\psi\omega)$. 
$Q^2$ resolution is estimated to be about $0.03$~(GeV/$c$)$^2$.
Used are the eight events shown in Fig.~\ref{fig:Q2-dist}.
  }
  \begin{tabular}{lrr}
  \hline
  \hline
  Item & Measurement & SBG model \\
  \hline
  Relative yield & $1.9 \pm 0.9$ & 1.0 \\
  $\langle Q^2 \rangle$ (GeV/$c$)$^2$ & $4.5\pm 0.7$ & 4.8 \\
  $\sqrt{\langle(Q^2 - \langle Q^2\rangle)^2 \rangle}$ (GeV/$c$)$^2$ 
	& $1.9\pm 0.8$ & 2.4 \\
  \hline
  \hline
  \end{tabular}
\end{table}

\section{\label{uncertainty}Systematic Uncertainties}

The largest uncertainty is associated with 
the $\pi^0$ selection efficiency, including the rate of fake $\pi^0$s.
By comparing the number of selected events using the different 
$\pi^0$ selection algorithms, we estimate 15\% uncertainty
associated with the $\pi^0$ selection algorithm.
Another uncertainty in $\pi^0$ detection is associated to 
fake $\pi^0$s from background photons.
In the data before applying $\pi^0$ selection, 
a significant number of low-energy photons, either true or fake, 
contaminate. 
These photons can produce fake $\pi^0$s.
To estimate the effect of such background photons, we look at variations in 
the ratio of events with identified $\pi^0$(s) to all events 
observed during the whole data-taking period. 
From this, we estimate a 5.6\% uncertainty 
after correcting the selection efficiency for the events with fake $\pi^0$.
This effect of background photons is also estimated using MC events 
simulated with different background conditions, which gives a 3\% variation. 
Conservatively, we use the larger 5.6\% as the systematic 
uncertainty in $\pi^0$ identification due to 
background photons.

Another large uncertainty is associated with $J/\psi$ 
identification.
The combined uncertainty in $J/\psi$ ID is 8\%. The largest
contribution, 7\%, to this comes from the difference in the ratio 
of the number of $J/\psi$ selected events, 
$N(J/\psi\rightarrow e^+e^-)/N(J/\psi\rightarrow \mu^+\mu^-)$, 
between the real data and MC. 
The other smaller contributions are 
the uncertainties in the efficiencies of electron and muon IDs, background
levels and radiative $\gamma$ corrections in the case of 
$J/\psi\rightarrow e^+e^-$ and the shapes of the invariant-mass 
distributions. 
They are estimated by the differences in the efficiencies 
between the real data and MC by varying the selection conditions.

The uncertainties in electron tagging, 5\%, 
and charged pion ID, 3\%, are estimated by 
the difference in the efficiencies between real data and MC 
by varying the selection conditions.
To calculate the detection efficiency, we set 
the fit region for selecting signal events. Because of 
the uncertainty in the $X(3915)$ distribution at or near the lower 
boundary of the fit region, 3.888~GeV/$c^2$, detection efficiency
will have an uncertainty, which is estimated to be 3\%.
The uncertainties in the $\omega$ selection, 2\%, and
the $p_T$-and-$E_{\rm obs}^*/E_{\rm exp}^*$ selection specified 
by Eq.~(\ref{eq:hyper}), 4\%, are estimated using MC by 
varying selection condition.
The uncertainty in the luminosity function, 
which is defined by Eq.~(\ref{eq:lum-func}), 3\%, is estimated 
from the uncertainties in QED modeling and numerical integration.
The other uncertainties are 2\% for missing $p_T$,
2\% for $||\phi({\rm tag})-\phi({\rm rest})|-\pi|$, 
1.8\% for track finding, 1.4\% for luminosity measurement, 
1\% for $p_T < 0.2$~GeV/$c$, 1\% for $Q^2$ numerical integration,
1\% for energy dependence in the detection efficiency, 
and 0.6\% for MC statistics.

Table~\ref{tab:systematics} lists a summary of systematic uncertainties.
As a total, combined quadratically, uncertainty in the reconstruction
efficiency is 20\%.
\begin{table}[ht]
  \centering
  \caption{\label{tab:systematics}Breakdown 
of contributions to the systematic uncertainty in the 
reconstruction efficiency.
  }
  \begin{tabular}{lr}
  \hline
  \hline
  Item & Uncertainty \\
  \hline
  $\pi^0$ selection algorithm		& 15\% \\
  $J/\psi$ ID				&  8\% \\
  Fake $\pi^0$ by background		& 5.6\% \\
  Electron tagging			&  5\% \\
  $p_T$-and-$E^*_{\rm obs}/E^*_{\rm exp}$ selection &  4\% \\
  Charged pion ID			&  3\% \\
  Luminosity function			&  3\% \\
  Efficiency window			&  3\% \\
  $\omega$ selection 			&  2\% \\
  Missing $p_T$				&  2\% \\
  $||\phi({\rm tag})-\phi({\rm rest})|-\pi|$ &  2\% \\
  Luminosity measurement		& 1.4\% \\
  Track finding				& 1.8\% \\
  $p_T < 0.2$~GeV/$c$			&  1\% \\
  $Q^2$ numerical integration		&  1\% \\
  Energy dependence in efficiency	&  1\% \\
  MC statistics				& 0.6\% \\
  \hline
  \textbf{Total} 			& \textbf{20\%} \\
  \hline
  \hline
  \end{tabular}
\end{table}

\section{\label{summary}Summary}

We performed the first measurement of the $Q^2$ distribution of $X(3915)$
production in single-tag two-photon interactions.
For signals, 
$7.9\pm 3.1({\rm stat.})\pm 1.5({\rm syst.})$ 
events are observed, 
while the expectation is $4.1\pm 0.7$,  
derived from the measured decay width at $Q^2=0$, 
$\Gamma_{\gamma\gamma}(0){\cal B}(X\rightarrow J/\psi\omega) = 54\pm 9$~eV, 
extrapolated to higher $Q^2$ region using the SBG $c\bar{c}$ model~\cite{SBG}. 
The shape of the $Q^2$ distribution is also
consistent with this model. These results can be used to
constrain non-$c\bar{c}$ models of the $X(3915)$ when
predictions for the $Q^2$ distribution become available.

\begin{acknowledgements}
This work, based on data collected using the Belle detector, which was
operated until June 2010, was supported by 
the Ministry of Education, Culture, Sports, Science, and
Technology (MEXT) of Japan, the Japan Society for the 
Promotion of Science (JSPS), and the Tau-Lepton Physics 
Research Center of Nagoya University; 
the Australian Research Council including grants
DP180102629, 
DP170102389, 
DP170102204, 
DE220100462, 
DP150103061, 
FT130100303; 
Austrian Federal Ministry of Education, Science and Research (FWF) and
FWF Austrian Science Fund No.~P~31361-N36;
the National Natural Science Foundation of China under Contracts
No.~11675166,  
No.~11705209;  
No.~11975076;  
No.~12135005;  
No.~12175041;  
No.~12161141008; 
Key Research Program of Frontier Sciences, Chinese Academy of Sciences (CAS), Grant No.~QYZDJ-SSW-SLH011; 
Project ZR2022JQ02 supported by Shandong Provincial Natural Science Foundation;
the Ministry of Education, Youth and Sports of the Czech
Republic under Contract No.~LTT17020;
the Czech Science Foundation Grant No. 22-18469S;
Horizon 2020 ERC Advanced Grant No. 884719, ERC Starting 
Grant No. 947006 "InterLeptons", and 
Grant No. 824093 "STRONG-2020" (European Union);
the Carl Zeiss Foundation, the Deutsche Forschungsgemeinschaft, the
Excellence Cluster Universe, and the VolkswagenStiftung;
the Department of Atomic Energy (Project Identification No. RTI 4002) and the Department of Science and Technology of India; 
the Istituto Nazionale di Fisica Nucleare of Italy; 
National Research Foundation (NRF) of Korea Grant
Nos.~2016R1\-D1A1B\-02012900, 2018R1\-A2B\-3003643,
2018R1\-A6A1A\-06024970, RS\-2022\-00197659,
2019R1\-I1A3A\-01058933, 2021R1\-A6A1A\-03043957,
2021R1\-F1A\-1060423, 2021R1\-F1A\-1064008, 2022R1\-A2C\-1003993;
Radiation Science Research Institute, Foreign Large-size Research Facility Application Supporting project, the Global Science Experimental Data Hub Center of the Korea Institute of Science and Technology Information and KREONET/GLORIAD;
the Polish Ministry of Science and Higher Education and 
the National Science Center;
the Ministry of Science and Higher Education of the Russian Federation, Agreement 14.W03.31.0026, 
and the HSE University Basic Research Program, Moscow; 
University of Tabuk research grants
S-1440-0321, S-0256-1438, and S-0280-1439 (Saudi Arabia);
the Slovenian Research Agency Grant Nos. J1-9124 and P1-0135;
Ikerbasque, Basque Foundation for Science, Spain;
the Swiss National Science Foundation; 
the Ministry of Education and the Ministry of Science and Technology of Taiwan;
and the United States Department of Energy and the National Science Foundation.
These acknowledgements are not to be interpreted as an endorsement of any
statement made by any of our institutes, funding agencies, governments, or
their representatives.
We thank the KEKB group for the excellent operation of the
accelerator; the KEK cryogenics group for the efficient
operation of the solenoid; and the KEK computer group and the Pacific Northwest National
Laboratory (PNNL) Environmental Molecular Sciences Laboratory (EMSL)
computing group for strong computing support; and the National
Institute of Informatics, and Science Information NETwork 6 (SINET6) for
valuable network support.

\end{acknowledgements}

\appendix

\section{DIFFERENTIAL CROSS SECTION}

The $Q^2$-differential $X(3915)$-production cross section in single-tag two-photon interactions 
is given by
\begin{equation}\label{eq:SBG-1}
\begin{array}{l}
{\displaystyle
\frac{d\sigma_{ee}(X(3915))}{dQ^2} = 
2\cdot 2\pi^2 \frac{(2J+1)\,2\Gamma_{\gamma\gamma}(0)}{M^2}
}
\\
{\displaystyle
\times
\left[f_{TT}(Q^2,M^2)\frac{d^2L^{TT}_{\gamma^*\gamma}}{dWdQ^2}\right.
}
\\
\multicolumn{1}{r}{\displaystyle
\left.\left.+ f_{LT}(Q^2,M^2)\frac{d^2L^{LT}_{\gamma^*\gamma}}{dWdQ^2}\right]
\right|_{W=M}
}
\end{array}
\end{equation}
where the factor 2 in the front stems 
from the two tag conditions ($e^-$-tag and $e^+$-tag), 
$J$ is the $X(3915)$ spin, $\Gamma_{\gamma\gamma}(0)$ is the $\gamma\gamma$
decay width of $X(3915)$ at $Q^2=0$,  $M$ is the mass of the $X(3915)$,
and 
$W$ is the energy of the two-photon system in its rest frame.
Furthermore, $f_{TT}(Q^2,M^2)$ and $f_{LT}(Q^2,M^2)$ are the
form factors for $X(3915)$ production in  
interactions of two transverse (virtual and quasireal) photons and of one longitudinal (virtual)
and one transverse (quasireal) photon, respectively;
$L^{TT}_{\gamma^*\gamma}$ as well as $L^{LT}_{\gamma^*\gamma}$ are 
the luminosity functions for the case of two transverse photons and 
for the case of one longitudinal and one transverse photon, respectively. 

Defining
\begin{eqnarray}
 \epsilon & = & 
{\displaystyle
\frac{L^{LT}_{\gamma^*\gamma}}{L^{TT}_{\gamma^*\gamma}}
}
\\
 L_{\gamma^*\gamma} & = &
{\displaystyle
L^{TT}_{\gamma^*\gamma} \label{eq:lum-func}
}
\end{eqnarray}
and 
\begin{equation}
{\displaystyle
f(Q^2,M^2,\epsilon) = f_{TT}(Q^2,M^2)+\epsilon f_{LT}(Q^2,M^2),
}
\end{equation}
Eq.~(\ref{eq:SBG-1}) can be rewritten as
\begin{equation}\label{eq:SBG-1-1}
\begin{array}{l}
{\displaystyle
\frac{d\sigma_{ee}(X)}{dQ^2} 
}   \\
{\displaystyle
=  8\pi^2 \frac{(2J+1)\Gamma_{\gamma\gamma}(0)}{M^2}
f(Q^2,M^2,\epsilon)\left.\frac{d^2L_{\gamma^*\gamma}}{dWdQ^2}
\right|_{W=M}.
} 
\end{array}
\end{equation}
We further introduce a $Q^2$-dependent decay function, 
\begin{equation}\label{eq:Zgamma}
{\displaystyle
Z_{\gamma^*\gamma}(Q^2,M^2,\epsilon)= \frac{f(Q^2,M^2,\epsilon) 
\Gamma_{\gamma\gamma}(0)}{(1+Q^2/M^2)} \, ,
}
\end{equation}
and rewrite 
Eq.~(\ref{eq:SBG-1-1}) as
\begin{equation}\label{eq:SBG-1-2}
\begin{array}{lll}
{\displaystyle
\frac{d\sigma_{ee}(X)}{dQ^2} 
} 
& = & 
{\displaystyle
8\pi^2 \frac{(2J+1)(1+Q^2/M^2)}{M^2}
Z_{\gamma^*\gamma}(Q^2,M^2,\epsilon)
} \\
& & 
{\displaystyle
\times\left.\frac{d^2L_{\gamma^*\gamma}}{dWdQ^2}\right|_{W=M}.}
\end{array}
\end{equation}
The differential event-yield distribution is
\begin{eqnarray}
{\displaystyle
 \frac{dN_{ee}(X)}{dQ^2} 
}
& = & 
{\displaystyle
\frac{d\sigma_{ee}(X)}{dQ^2}\varepsilon_{\rm eff}(Q^2)
    L_{\rm int}{\cal B}(X\rightarrow J/\psi\omega)  \nonumber 
}
\\
  & &
{\displaystyle
\times\,{\cal B}(J/\psi\rightarrow \ell^+\ell^-)
	{\cal B}(\omega\rightarrow \pi^+\pi^-\pi^0) \, , 
}
\\
  & = &
{\displaystyle
\, 8\pi^2 \frac{(2J+1)(1+Q^2/M^2)}{M^2}  
  Z_{\gamma^*\gamma}(Q^2,M^2,\epsilon) \nonumber 
}
\\
  & & 
{\displaystyle
\times \, \varepsilon_{\rm eff}(Q^2) L_{\rm int} 
          \left.\frac{d^2L_{\gamma^*\gamma}}{dWdQ^2}\right|_{W=M} 
	\nonumber 
}
\\  
  & & 
{\displaystyle
\times \, {\cal B}(X\rightarrow J/\psi \omega) \, 
	{\cal B}(J/\psi\rightarrow \ell^+\ell^-) \nonumber 
}
\\
  & & 
{\displaystyle
\times \, {\cal B}(\omega\rightarrow \pi^+\pi^-\pi^0) \, , \label{eq:SBG-2}
}
\end{eqnarray}
where $L_{\rm int}$ is the integrated luminosity, 
${\cal B}(J/\psi\rightarrow \ell^+\ell^-)$ is the branching fraction of
$J/\psi$ decaying to either an electron pair or a muon pair, 
${\cal B}(\omega\rightarrow\pi^+\pi^-\pi^0)$ is the branching
fraction of $\omega$ decaying to three pions.
Rearranging Eq.~(\ref{eq:SBG-2}), one can 
relate \(Z_{\gamma^*\gamma}(Q^2,M^2,\epsilon)\, 
{\cal B}(X\rightarrow J/\psi \omega)\)
to the event-yield distribution:
\begin{equation}\label{eq:C}
  Z_{\gamma^*\gamma}(Q^2,M^2,\epsilon)
	{\cal B}(X\rightarrow J/\psi \omega)
  = C(Q^2,M^2)\frac{dN_{ee}(X)}{dQ^2} 
\end{equation}
with
\begin{equation}
\begin{array}{l}
{\displaystyle
 1/C(Q^2,M^2)} \\
{\displaystyle
 = 8\pi^2 \frac{(2J+1)(1+Q^2/M^2)}{M^2}  
	\left.\frac{d^2L_{\gamma^*\gamma}}{dWdQ^2}\right|_{W=M}} \\
{\displaystyle
  \ \ \ \times\, \varepsilon_{\rm eff}(Q^2)  \, L_{\rm int} \, 
	{\cal B}(J/\psi\rightarrow \ell^+\ell^-) \, 
	{\cal B}(\omega\rightarrow \pi^+\pi^-\pi^0) \, . }\label{Coeff}
\end{array}
\end{equation}

For the production of $J^P = 0^+$ particles, as $X(3915)$, 
the $f_{LT}$ component does not contribute and hence the  
$\epsilon$ dependence of \(Z_{\gamma^*\gamma}\) drops out. 
Furthermore, with $J=0$ and using the integrated luminosity in this analysis, 
$L_{\rm int} = 825~\text{fb}^{-1}$, as well as  
${\cal B}(J/\psi\rightarrow\ell^+\ell^-)=0.11932$ and
${\cal B}(\omega\rightarrow\pi^+\pi^-\pi^0)=0.892$~\cite{PDG-2}, 
Eq.~\eqref{Coeff} simplifies to
\begin{equation}
\begin{array}{lcl}
 1/C(Q^2,M^2) & = & 
  {\displaystyle 8\pi^2 \frac{1+Q^2/M^2}{M^2} 
  }
 \\
  & & {\displaystyle \times \, 3.418\times 10^{13} 
	\varepsilon_{\rm eff}(Q^2) \!  
	\left.\frac{d^2L_{\gamma^*\gamma}}{dWdQ^2}\right|_{W=M} \!\! .
  }
\end{array}
\end{equation}

\begin{figure}[ht]
  \centering
  \includegraphics*[width=0.40\textwidth]{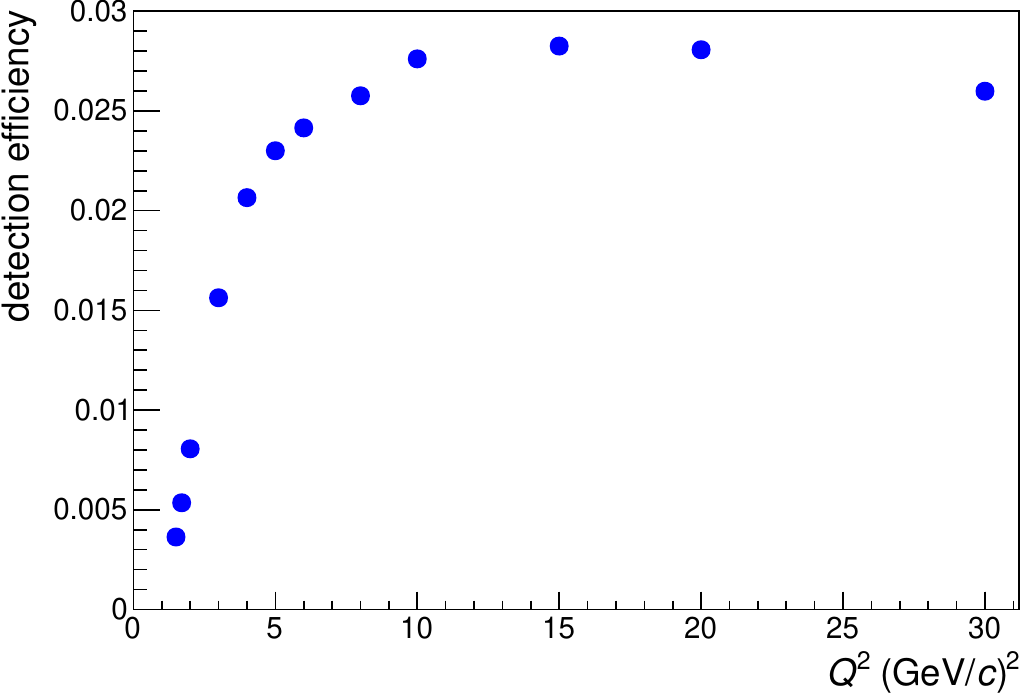}
  \caption{\label{fig:eff}Detection efficiency as a function of $Q^2$ 
as obtained from a MC simulation.
}
\end{figure}

\begin{figure}[hb]
  \centering
  \includegraphics*[width=0.40\textwidth]{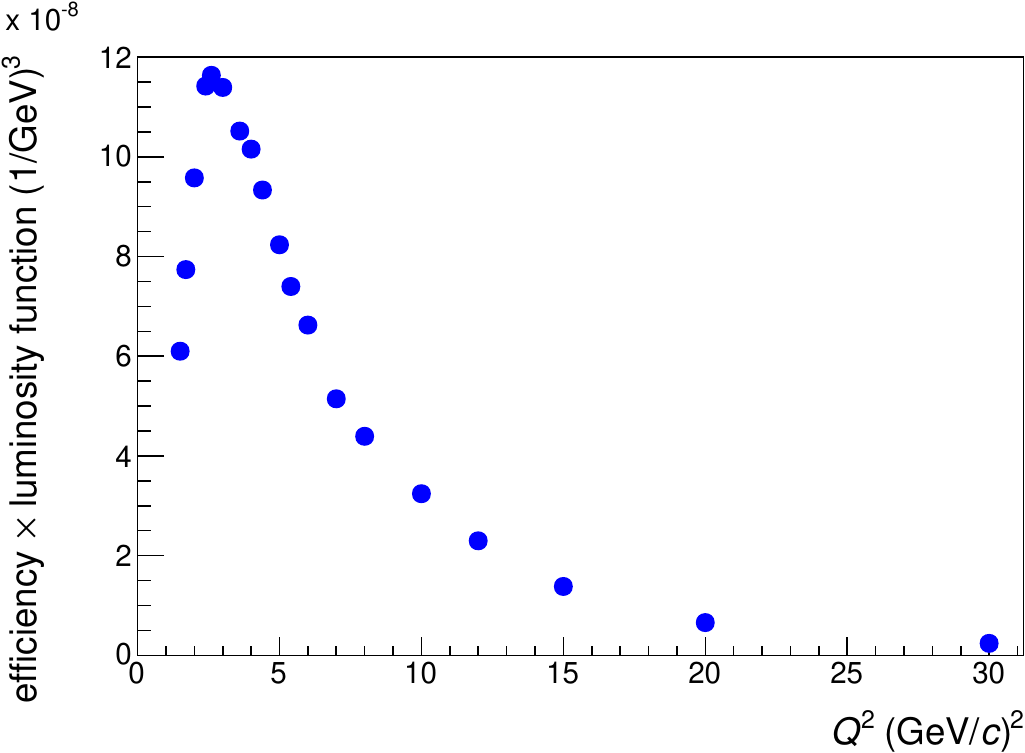}
  \caption{\label{fig:eff-lum}
Detection efficiency times luminosity function as a function of $Q^2$.
Ordinate is $\left.\varepsilon_{\rm eff}(Q^2)\frac{d^2L_{\gamma^*\gamma}}
{dWdQ^2}\right|_{W=M}$~(1/GeV)$^3$.}
\end{figure}

In order to obtain numerical values for $C(Q^2,M^2)$, the
detection efficiency is calculated using MC events. Figure~\ref{fig:eff}
shows the resulting efficiency as a function of $Q^2$.
The product of the efficiency and the luminosity function is 
presented in Figure~\ref{fig:eff-lum}.
This distribution shows our sensitivity for measuring the 
$Q^2$ distribution; the sensitive
region is between $Q^2=1.5$~(GeV/$c$)$^2$ and $Q^2\approx 10$~(GeV/$c$)$^2$.
Finally, numerical values for $C(Q^2,M^2)$ for $M=3.918$~GeV/$c^2$ are  
listed in Table~\ref{tab:CQ2}.

\begin{table}[ht]
  \centering
  \caption{\label{tab:CQ2}$C(Q^2,M^2)$: conversion factor from 
the number of events to 
$Z_{\gamma^*\gamma}(Q^2/M^2){\cal B}(X\rightarrow J/\psi\omega)$
as a function of $Q^2$.}
  \begin{footnotesize}
  \begin{tabular}{rrrrrrrr}
    \hline
    \hline
    $Q^2$~(GeV/$c$)$^2$ &
 1.5 & 1.7 & 2.0 & 2.4 & 2.6 & 3.0 & 3.6 \\
    \hline
    $C(Q^2,M^2)$ $\times 10^{-8}$ & 
	 8.49 & 6.62 & 5.25 & 4.31 & 4.18 & 4.18 & 4.38 \\
    \hline
    \hline
    $Q^2$~(GeV/$c$)$^2$ &
 4.0 & 4.4 & 5.0 & 5.4 & 6.0 & 7.0 & 8.0 \\
    \hline
    $C(Q^2,M^2)$ $\times 10^{-8}$ & 
	 4.44 & 4.74 & 5.21 & 5.69 & 6.17 & 7.59 & 8.51 \\
    \hline
    \hline
    $Q^2$~(GeV/$c$)$^2$ &
 10.0 & 12.0 & 15.0 & 20.0 & 30.0 & & \\
    \hline
    $C(Q^2,M^2)$ $\times 10^{-8}$ & 
	10.62 & 13.90 & 20.82 & 37.58 & 80.09 & & \\
    \hline
    \hline
  \end{tabular}
  \end{footnotesize}
\end{table}

The theoretical expression for the decay function $Z_{\gamma^*\gamma}$ is given in
the SBG model~\cite{SBG} as
\begin{equation}\label{eq:SBG-0+}
  Z_{\gamma^*\gamma}(Q^2/M^2) =  
  \frac{1}{(1 + Q^2/M^2)^4}\left(1+\frac{Q^2}{3M^2}\right)^2
  \Gamma_{\gamma\gamma}(0)
\end{equation}
for $J^P = 0^+$, while it is  
\begin{equation}\label{eq:SBG-2+}
\begin{array}{l}
{\displaystyle
  Z_{\gamma^*\gamma}(Q^2/M^2,\epsilon)
} \\
{\displaystyle
 =  \frac{1}{(1 + Q^2/M^2)^4}\left(1+\frac{Q^4}{6M^4}
+\epsilon\frac{Q^2}{M^2}\right)
  \Gamma_{\gamma\gamma}(0)
}
\end{array}
\end{equation}
in case of $J^P = 2^+$.

\bibliography{ref.bib}

\end{document}